\documentclass[conference]{IEEEtran}

\usepackage{xcolor}
\usepackage{enumitem}
\usepackage{subfigure}
\usepackage[linesnumbered,ruled,vlined]{algorithm2e}
\SetEndCharOfAlgoLine{}

\IEEEoverridecommandlockouts
\usepackage{cite}
\usepackage{amsmath,amssymb,amsfonts}
\usepackage{algorithmic}
\usepackage{graphicx}
\usepackage{textcomp}
\usepackage{xcolor}
\def\BibTeX{{\rm B\kern-.05em{\sc i\kern-.025em b}\kern-.08em
    T\kern-.1667em\lower.7ex\hbox{E}\kern-.125emX}}
\begin{document}

\title{
SAGkit: A Python SAG Toolkit for Response\\Time Analysis of Hybrid-Triggered Jobs
}

\author{
    \IEEEauthorblockN{Ruide Cao$^{\dag1,5}$, Zhuyun Qi$^{\dag2,4}$, Qinyang He$^{3}$, Chenxi Ling$^{1,4}$, Yi Wang$^{*1,4}$, Guoming Tang$^{5}$}
    \IEEEauthorblockA{$^1$Institute of Future Networks, SUSTech\ $^2$Tsinghua SIGS, Tsinghua University\ $^3$Nankai University}
    \IEEEauthorblockA{$^4$Peng Cheng Laboratory\ $^5$The Hong Kong University of Science and Technology (Guangzhou)}
}

\maketitle

\renewcommand{\thefootnote}{\fnsymbol{footnote}}
\footnotetext{$^\dag$Co-first authors: Ruide Cao and Zhuyun Qi.}
\footnotetext{$^*$Corresponding author: Yi Wang (wy@ieee.org). This work is supported by the Guangdong High-Level Talents Special Support Program (2021TX05X205), the Guangdong Basic and Applied Basic Research Foundation (2024A0101010001), and Peng Cheng Laboratory The Major Key Project of PCL (PCL2023A03).}

\begin{abstract}
For distributed control systems, modern latency-critical applications are increasingly demanding real-time guarantees and robustness. Response-time analysis (RTA) is useful for this purpose, as it helps analyze and guarantee timing bounds. However, conventional RTA methods struggle with the state-space explosion problem, especially in non-preemptive systems with release jitter and execution time variations. In this paper, we introduce SAGkit, a Python toolkit that implements the schedule-abstraction graph (SAG) framework. SAGkit novelly enables exact and sustainable RTA of hybrid-triggered jobs by allowing job absence on the SAG basis. Our experiments demonstrate that SAGkit achieves exactness with acceptable runtime and memory overhead. This lightweight toolkit empowers researchers to analyze complex distributed control systems and is open-access for further development.
\end{abstract}


\section{Introduction}
Under the Industry 4.0 trend, IT/OT convergence is advancing toward increasingly diverse and complex application scenarios \cite{de2023single}. Theoretical guarantees that the distributed cyber-physical systems are safe and reliable help to achieve predictable and robust physical control \cite{girault2001fault, hua2014cooperative, wang2019frame}. Along with functional correctness guarantees, timing correctness is equally important, as late `correct' behavior can be harmful \cite{krotofil2014cps}.

Response-time analysis (RTA) techniques are extensively applied to distributed real-time control systems \cite{gu2003integrated, soni2024hybrid}. By calculating the maximum time it takes for each task to complete after its release, RTA could provide an upper bound on all possible response times, which is essential for guaranteeing timing correctness. This theoretical guarantee allows system designers to assess whether critical tasks could be finished within the required time, even in the presence of release jitter and execution time variations. However, conventional RTA tools \cite{burmyakov2015exact, sun2016pre} are not scalable enough to support analyzing the increasingly complex modern distributed control systems.

A recent exact and sustainable RTA framework named schedule-abstraction graph (SAG) \cite{nasri2017exact} scales well for non-preemptive situations in both processor and network contexts. As a promising approach, SAG has been extended to analyze the CAN message schedules \cite{nasri2018using} and the time-aware shapers in time-sensitive networking \cite{srinivasan2021work}. More recently, a dynamic job set enabled SAG is proposed \cite{gohari2023response}. It enables the emerging jobs to be dynamically taken into account during the construction of SAG, thereby enhancing fault tolerance.

These SAG-based approaches allow the execution time of jobs to vary over a range, but all assume this range to be continuous. Such an assumption prevents them from being applied to hybrid event-triggered control (HETC) systems, where time-triggered and event-triggered mechanisms are combined to make the systems more adaptive and efficient. Common HETC systems include automotive and aerospace systems \cite{bai2020autoe2e, liao2023hybrid, liang2023hybrid} where embedded real-time control is performed and resource efficiency is critical. In these systems, scheduled jobs might be absent due to specific event conditions not being met \cite{fei2021discontinuous, zhang2021hybrid}. To further improve its applicability, the SAG framework must integrate job absence compatibility.

Hybrid-triggered (HT) jobs refer to jobs that are primarily scheduled at predetermined times but are executed when specific conditions or events are met. This dual-trigger mechanism incorporates the predictability of time-triggered systems with the responsiveness of event-triggered systems. Unlike purely time-triggered jobs, which execute strictly at pre-defined times, or traditional event-triggered jobs, which activate only in response to external signals, HT jobs offer more flexibility to real-time and cyber-physical systems. By incorporating conditional execution within a time-based framework, HT jobs allow systems to optimize job scheduling dynamically, balancing determinism with conditional responsiveness. The HETC model is especially suited for applications where jobs must be available at regular intervals but do not always need to be executed, thus conserving computational and communication resources while maintaining time-sensitive operations.

We build SAGkit on our previous work of \cite{cao2024extending} to push the applicability limitations of existing SAG-based RTA tools. The contributions of this work are summarized as follows:
\begin{itemize}[leftmargin=20pt]
\item We extend the original SAG to a hybrid version, which is capable of supporting exact and sustainable RTA for HT jobs. A variant of SAG named \textit{Hybrid SAG} is proposed, allowing jobs to have not only release time jitter and execution time variations but also absence cases.

\item We conduct experiments to validate the effectiveness of \textit{Hybrid SAG} and to measure its overhead. The results demonstrate the exactness and scalability of our method, where the increases in both time and space overhead are considered sufficiently acceptable.

\item We package our SAG constructing and analyzing code into a Python toolkit named \textit{SAGkit} for open-source releases\footnote{Releases are available at https://github.com/RyderCRD/sagkit.} and distribution\footnote{https://pypi.org/project/sagkit. Install with '\texttt{pip install sagkit}'.}. One can easily reproduce all experimental results in this paper with SAGkit. It can also be used as a basis for subsequent studies by other researchers.

\end{itemize}

\noindent \textbf{Related tools.} To the best of our knowledge, there are no SAG-based tools that are compatible with HT jobs without losing exactness. However, there are two useful tools that can be used to construct SAGs: (I) the official repository of SAG (in C++) \cite{nelissen2017sag}; (II) Schedule-Abstraction Graph in GO \cite{porya2022sag-go}. Both of them are on the basis that the execution variation is continuous, and thus, neither of them natively supports the job absence. In addition, prior to SAGkit, there was no publicly available SAG constructor in Python.

\section{System Model}

In this toolkit, the input for constructing SAG is defined as individual jobs rather than tasks. While some theoretical scheduling frameworks distinguish between tasks (repeating units that generate multiple jobs over time) and jobs (specific instances of work with distinct release times and deadlines), the concept of "task" does not directly impact the scheduling process. We treat each job as an independent unit based on its individual attributes---additional complexity of task-level abstraction is thus avoided. For situations where tasks exist, they can be analyzed by breaking down into jobs \cite{nasri2017exact}.

\subsection{Jobs and Schedulers}
Consider scheduling a finite set of non-preemptive jobs $J$ on a uniprocessor platform, where each job $J_i$ can be characterized as $([r_i^{min}, r_i^{max}], [C_i^{min}, C_i^{max}], d_i, p_i, E_i])$. With $[r_i^{min}, r_i^{max}]$ being the release/arrival time jitter, the execution time variation is denoted by $[C_i^{min}, C_i^{max}]$. A job $J_i$ may arrive at any time $r_i$ that $r_i^{min} \le r_i \le  r_i^{max}$ and be finished at any time cost $C_i$ that $C_i^{min} \le C_i \le C_i^{max}$. A smaller $p_i$ value implies higher priority, and $d_i$ indicates the absolute deadline. On top of the original SAG model \cite{nasri2017exact}, we add a 0-1 variable $E_i$ to indicate whether $J_i$ has external event trigger conditions, i.e., whether it is an HT job.

Despite the jitter in arrivals, a completely time-triggered job is certain to arrive in every hyperperiod. In contrast, an HT job may be absent at runtime because some event conditions (other than time) are not met. When a job $J_i$ is absent, we treat its actual execution time $C_i$ as 0 since it does not require any resources to actually get executed.

Multiple scheduling algorithms are natively supported in the SAG framework, including non-preemptive fixed-priority (NP-FP) \cite{tindell1994extendible}, non-preemptive earliest deadline first (NP-EDF) \cite{jeffay1991non}, and a more recent proposed precautions rate-monotonic \cite{nasri2014precautious}. When several jobs have been released and are waiting to be executed, these scheduling methods could be applied to decide which takes precedence. For clarity, we default to using NP-FP as the scheduler in the rest of this paper.

\subsection{Execution Scenarios}

We use $\{.\}$ to denote a set where the order is irrelevant and $\langle.\rangle$ for a sequence of items. A job set $J$ is considered schedulable under a given scheduling algorithm if every possible execution scenario for $J$ meets all deadlines. The term \textit{execution scenario} is defined as

\noindent \textbf{Definition 1.} An \textit{execution scenario} $\gamma=(C, R)$ for a certain set of jobs $J=\{J_1, J_2, \dots, J_m\}$ is a sequence of execution times $C=\langle C_1, C_2, \dots, C_m\rangle$ and release times $R=\langle r_1, r_2, \dots, r_m \rangle$ such that, $\forall J_i \in J, r_i \in [r_i^{min}, r_i^{max}]$ and
\begin{equation*}
C_i \in \left\{
\begin{array}{ll} 
[C_i^{\min}, C_i^{\max}], & E_i = 0, \\ 
\{0\} \cup [C_i^{\min}, C_i^{\max}], & \text{otherwise.}  
\end{array}
\right.
\end{equation*}

Differing from the definition of \textit{execution scenario} in \cite{nasri2017exact}, the support for HT jobs is included in this definition. When $J_i$ is a probable absentee job, an execution scenario of $C_i = 0$ is separately considered. Based on the definition of an execution scenario, we refer to \cite{ranjha2022partial} to give the definition of an \textit{exact} RTA as

\noindent \textbf{Definition 2.} An analysis is an \textit{exact} RTA iff (I) there is no execution scenario such that any job has a response time smaller than the earliest response time or larger than the latest response time returned by the analysis for that job, and (II) for each job, there must be an execution scenario where the job experiences the computed earliest response time and another execution scenario where it experiences the computed latest response time.

With Def. 1 and Def. 2, we then present our hybrid SAG that can remain exact on the HT job sets in the next section.

\section{Hybrid SAG}
All possible execution scenarios can be expressed within a SAG. Formally, a SAG is a \textit{directed acyclic graph} where the label of each edge is a job $J_x \in J$, and the label of each vertex is a state $S_i \in S$ with a reachable time interval $[e_i, l_i]$. The $e_i$ and $l_i$ represent the earliest and latest finish time of the job sequence in any path that connects the root vertex $S_1$ to $S_i$. Except for $S_1$, there exists at least one execution scenario for any $S_i$ and any time $t \in [e_i, l_i]$ that the last job in one of the corresponding paths from $S_1$ to $S_i$ finishes by time $t$. Also, every possible job sequence has its own branch (i.e., a path originating at the root $S_1$) in this graph.

To present the basic idea, we use SAG to analyze the response time of a smoke alarm consisting of a smoke sensor and an alarm. Assume the sensor detects smoke periodically, which takes 1 or 2 time units, an alert job could be released to the alarm between 1 and 2 in each period. In case an alert action takes 3 to 4 time units to complete, the following SAGs can be generated:

\begin{figure}[htbp]
    \centering
    \begin{minipage}[t]{0.23\textwidth}
        \centering
        \includegraphics[scale=0.1]{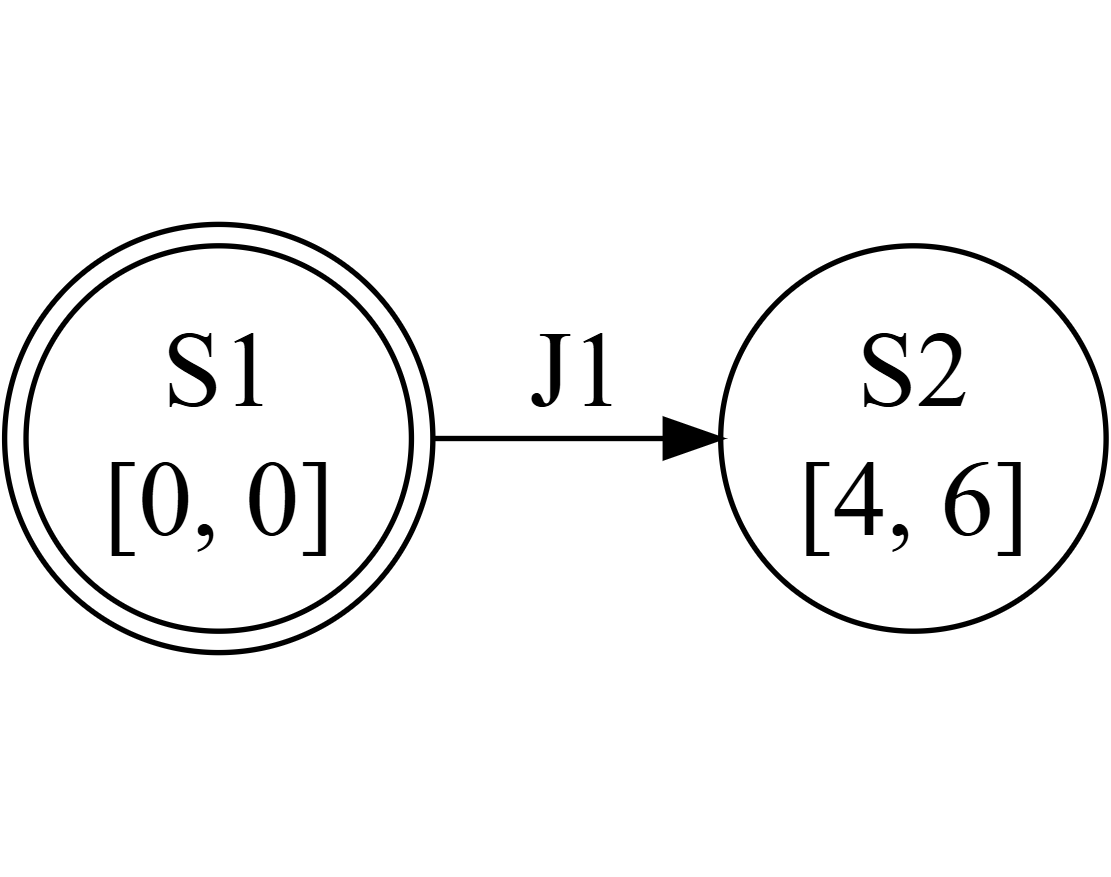}
        \caption{Original SAG.}
        \label{fig_simple_1}
    \end{minipage}
    \begin{minipage}[t]{0.23\textwidth}
        \centering
        \includegraphics[scale=0.1]{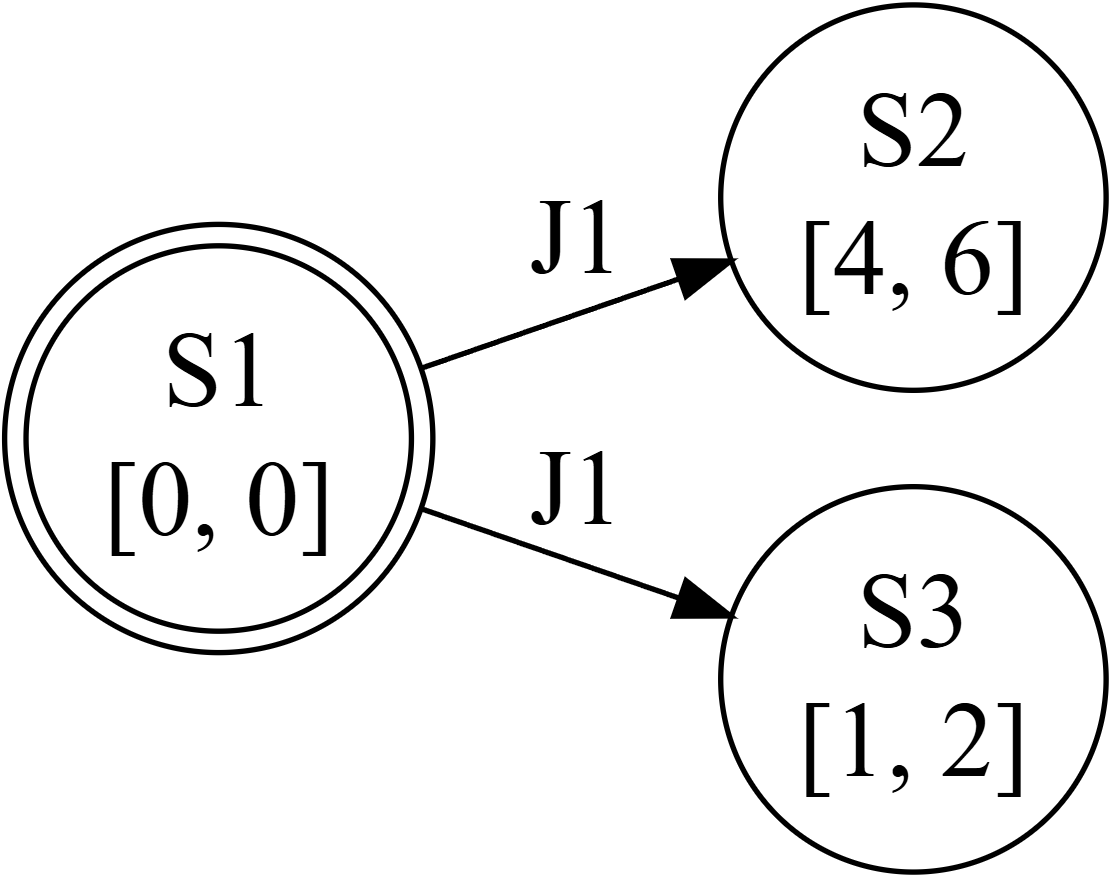}
        \caption{Hybrid SAG.}
        \label{fig_simple_2}
    \end{minipage}
\end{figure}
\begin{figure}[tb]
    \centering
    \subfigure[Original SAG]{\includegraphics[width=0.8\columnwidth]{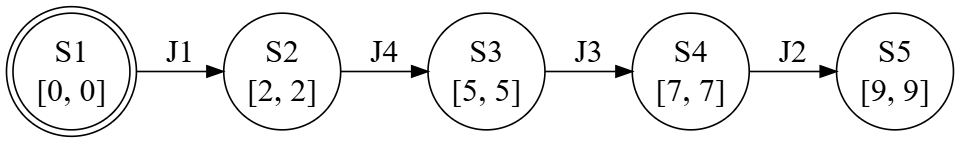}}
    \subfigure[Extended SAG]{\includegraphics[width=0.8\columnwidth]{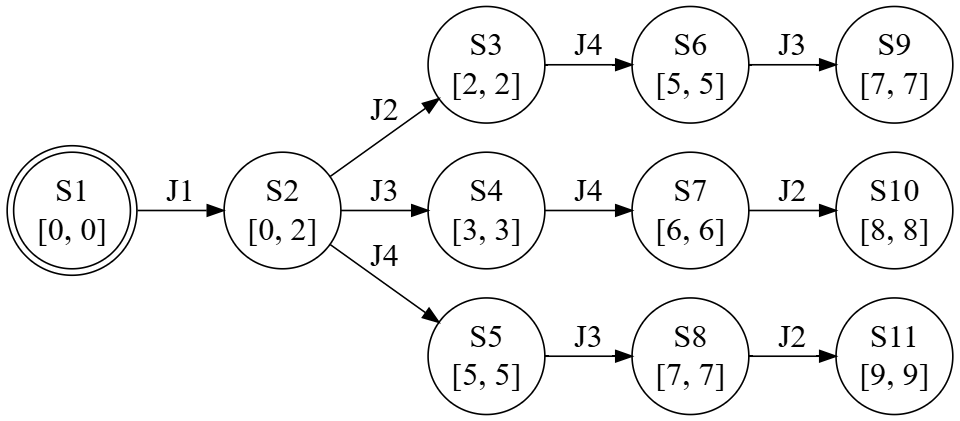}}
    \subfigure[Hybrid SAG (proposed)]{\includegraphics[width=0.8\columnwidth]{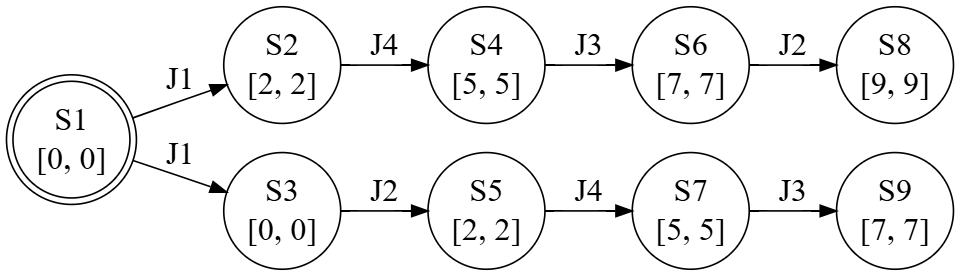}}
    \caption{A plain example where all jobs are free of release jitter and execution time variations.}
    \label{fig_example_1}
\end{figure}

Modeling the alert job as $J_1 = ([1, 2], [3, 4], 6, 1, 1])$, the original SAG analyzed the alarm response time as 4 to 6, as Fig. \ref{fig_simple_1} shows. Since $r_1^{min} = 1$ and $C_1^{min} = 3$, the job $J1$ is released earliest at moment 1 and takes a minimum of 3 units of time to finish. In this best case, the alarm action does respond in 4 time units. Similarly, in the worst case ($r_1 = 2$, $C_1 = 4$), the response time of the alarm could reach 6. However, not every detection should be followed by an alarm. If the smoke concentration does not reach a specific threshold, an alarm is not expected, which is the case for most times. Our proposed hybrid SAG is as shown in Fig. \ref{fig_simple_2}. When the sensing signal arrives between 1 and 2 time units without triggering an alarm, the scenarios are analyzed as the $S3$ state.

None of the states with $J1$ completed ($S2, S3$) had a maximum reachable time greater than the absolute deadline $d_1 = 6$. Both the original SAG and the hybrid SAG, although different, were able to validate this smoke alarm as safe. We next demonstrate the motivation and significance of proposing the hybrid SAG through two more complex examples.

\subsection{Examples}

\begin{figure}[tb]
    \centering
    \subfigure[Original SAG]{\includegraphics[width=0.8\columnwidth]{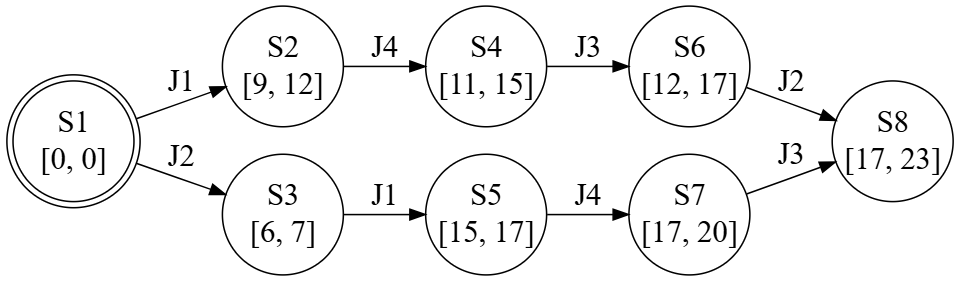}}
    \subfigure[Extended SAG]{\includegraphics[width=0.8\columnwidth]{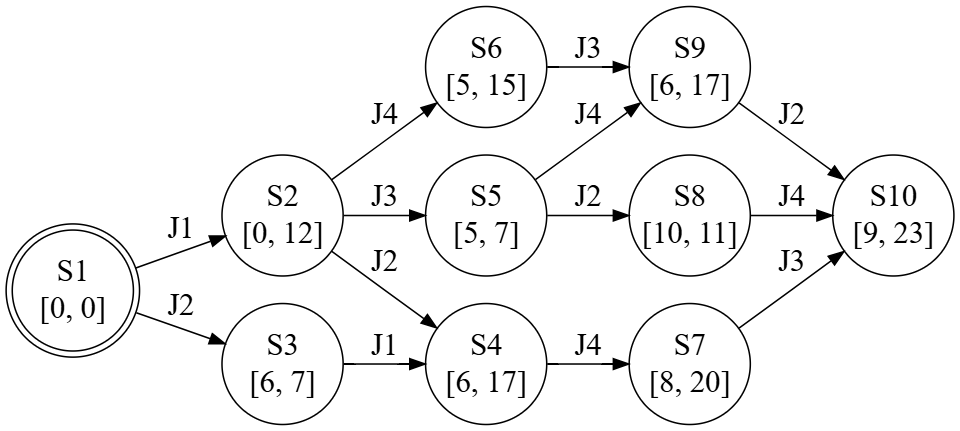}}
    \subfigure[Hybrid SAG (proposed)]{\includegraphics[width=0.8\columnwidth]{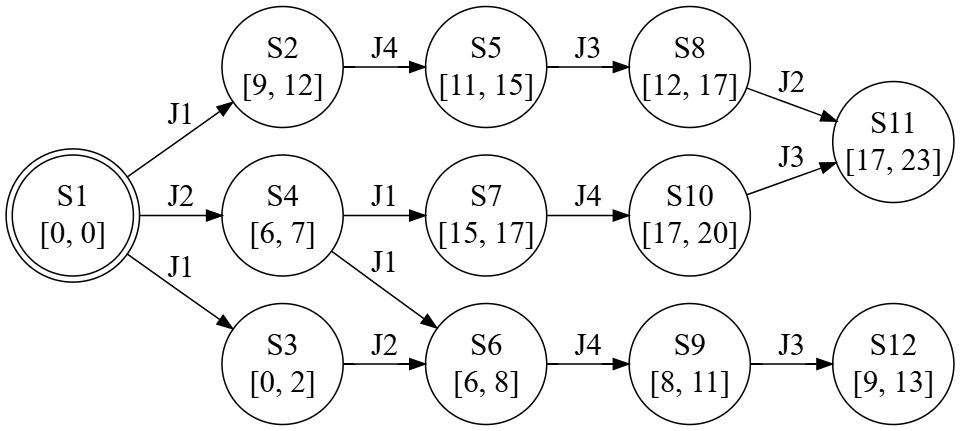}}
    \caption{A more realistic example where all jobs have release jitter and execution time variations.}
    \label{fig_example_2}
\end{figure}

We constructed two examples as shown in Fig. \ref{fig_example_1} and Fig. \ref{fig_example_2}. The graphs from the same figure were constructed based on the same set of jobs (according to Table \ref{jobsets_table}) using different SAG construction algorithms. In addition to the \textbf{original SAG}, which could not analyze job absences, we also made an intuitive variant to take the absences into account. This variant is called \textbf{extended SAG}, and it sets the minimum execution time of all HT jobs directly to 0. We then compare the proposed \textbf{hybrid SAG} with these two SAGs.

\begin{table}[b]
    \caption{Parameters of Example Job Sets}
    \centering
    \scalebox{0.88}{
        \begin{tabular}{cccccc|cccccc}
        \hline
        \multicolumn{6}{c|}{Job Set of Example 1}            & \multicolumn{6}{c}{Job Set of Example 2}              \\
        i & $r_i$      & $C_i$      & $d_i$ & $p_i$ & $E_i$ & i & $r_i$      & $C_i$       & $d_i$ & $p_i$ & $E_i$ \\ \hline
        1 & {[}0, 0{]} & {[}2, 2{]} & 5     & 1     & 1     & 1 & {[}0, 2{]} & {[}9, 10{]} & 20    & 1     & 1     \\
        2 & {[}0, 0{]} & {[}2, 2{]} & 10    & 4     & 0     & 2 & {[}1, 2{]} & {[}5, 6{]}  & 25    & 4     & 0     \\
        3 & {[}1, 1{]} & {[}2, 2{]} & 10    & 3     & 0     & 3 & {[}4, 5{]} & {[}1, 2{]}  & 25    & 3     & 0     \\
        4 & {[}2, 2{]} & {[}3, 3{]} & 5     & 2     & 0     & 4 & {[}3, 6{]} & {[}2, 3{]}  & 25    & 2     & 0     \\ \hline
        \end{tabular}
    }
    \label{jobsets_table}
\end{table}

\textit{Example 1.} There are 4 jobs in the job set without any release time jitter or execution time variation. The original SAG only has one path from the root to the leaf, as Fig. \ref{fig_example_1} (a) shows, indicating the only job ordering $\langle J_1, J_4, J_3, J_2 \rangle$. However, $J_1$ might be absent at runtime, which could lead to another job ordering $\langle J_1, J_2, J_4, J_3 \rangle$. Although $J_4$ has a higher priority and will be prioritized by the FP scheduler, $J_2$ is ahead of $J_4$ in this ordering because of its earlier arrival. 

While the original SAG is under-considering, the extended SAG over-considered an unreachable execution scenario, along with a job order $\langle J_1, J_3, J_4, J_2 \rangle$ that would not actually occur at runtime. The reason for this situation is that when we reduce the minimum execution time $C_1^{min}$ from 2 to 0, additional unreachable execution scenarios of $C_1 = 1$ were introduced. In fact, $C_1 = 1$ (i.e., $J_1$ finished in 1 time unit) never happens.

\textit{Example 2.} There are 4 jobs having both release jitter and execution time variations. Depending on which arrives first, both $J_1$ and $J_2$ may be the first job to be executed. While the original SAG and the extended SAG only create a single vertex $S_2$ for the $J_1$ executed state, the hybrid SAG separately creates two vertices corresponding to cases where $J_1$ is executed and is absent. As a consequence, the following states are all affected and vary.

The earliest response time of $J_3$ proves the significance of our proposed hybrid SAG. In the original SAG (Fig. \ref{fig_example_2} (a)), $J_3$ was first finished in state $S_6$, and its earliest response time was pessimistically analyzed as 12. In the execution scenario where $J_3$ is completed at the earliest, $J_1$ is released at 0, executed with 9 time units, and finished at 9. Then, Job $J_4$ took 2 units of time to finish until 11, and $J_3$ itself took another 1 time unit till 12. Conversely, in the extended SAG (Fig. \ref{fig_example_2} (b)), a state $S_5$ that does not actually occur prematurely analyzed the earliest response time of $J_3$ as 5. In fact, $J_3$'s earliest response time is 9, as exactly analyzed by the hybrid SAG shown in Fig. \ref{fig_example_2} (c).

\section{SAGkit Overview}

We implement the above system model along with the three different SAG construction algorithms in Python3, relying on Graphviz \cite{ellson2002graphviz} for graph visualization. We package the code into a toolkit named SAGkit, and the core components are described as follows:

\begin{itemize}[leftmargin=10pt]
\item \textsf{job.py} defines the \textsf{Job} class, encapsulating essential attributes related to individual jobs in the SAG framework, including release time jitter interval, best-case and worst-case execution time, deadline, priority, and other job-specific details. This module forms a foundational component for representing and manipulating jobs consistently across various scheduling and construction algorithms.
\item \textsf{state.py} defines the \textsf{State} class, representing the different execution states of jobs within the SAG. Each state, as a vertex in the graph, is associated with a set of jobs and other states. This module is essential for tracking execution progress and constructing SAGs to represent possible execution paths and dependencies through connection status.
\item \textsf{jobset\_generator.py} generates job sets with customizable parameters, including the event-triggered ratio, utilization, and number of jobs, and saves them to files.
\item \textsf{fp\_scheduler.py} implements the FP scheduling algorithm, a commonly used scheduling policy where jobs are scheduled based on their priorities.
\item \textsf{original\_constructor.py} implements the original algorithm for constructing a SAG \cite{nasri2017exact} in its \textsf{Constructor} class. A constructor object reads job attributes from the generated job set files and performs the construction of SAG based on a given scheduling algorithm.
\item \textsf{extended\_constructor.py} inherits \textsf{Constructor} class and introduces support for HT jobs in an intuitive way, i.e., extending the minimum execution time of all potentially absent jobs to 0. In this way, it ensures that absent execution scenarios are analyzed, thus guaranteeing the reliability of the analysis and the security of the target system.
\item \textsf{hybrid\_constructor.py} also inherits \textsf{Constructor} class and supports HT jobs, but in a manner that does not compromise exactness, i.e., it splits every absent execution scenario into a separate state to be analyzed.
\end{itemize}

\section{Evaluation}
Hybrid SAG maintains exactness over the HT job sets, thus enabling correct RTA on the HETC systems. However, there is a potential for exponential state explosion due to the separate consideration of HT jobs. To investigate the performance of the hybrid SAG in terms of both \textbf{exactness} and \textbf{scalability}, a series of experiments were conducted with SAGkit.

\subsection{Experiment Setup}

According to the characteristics of a real-world Bosch automotive engine control application provided by Kramer et al. \cite{kramer2015real}, the typical total number of jobs ranges from 1,000 to 1,500. Of these jobs, about 85\% are strictly time-triggered, accompanied by about 15\% synchronized to the rotation angle of the internal crankshaft. We consider this 15\% jobs to be HT because of their hybrid event-triggered nature.

In experiments, each job set contains 1000 randomly generated jobs, with $r_i^{min} \in [1, 9990]$, $r_i^{max} \in [r_i^{min}+0, r_i^{min}+9]$, $C_i^{min} \in [2, C_U]$, $C_i^{max} \in [C_i^{min}+1, C_i^{min}+4]$, $d_i = 9999$, $p_i \in [1, 10]$ for each job $J_i$. The desired utilization was reached by controlling the value of $C_U$, which equals $(U / 5 - 7)$ for $U \in [45, 75]$. An Intel i7-12700KF processor clocked at 3.60 GHz and 32 GB memory were used for running SAGkit.

\subsection{Exactness}

\begin{figure*}[t]
    \centering
    \subfigure[Original SAG]{\includegraphics[scale=0.20]{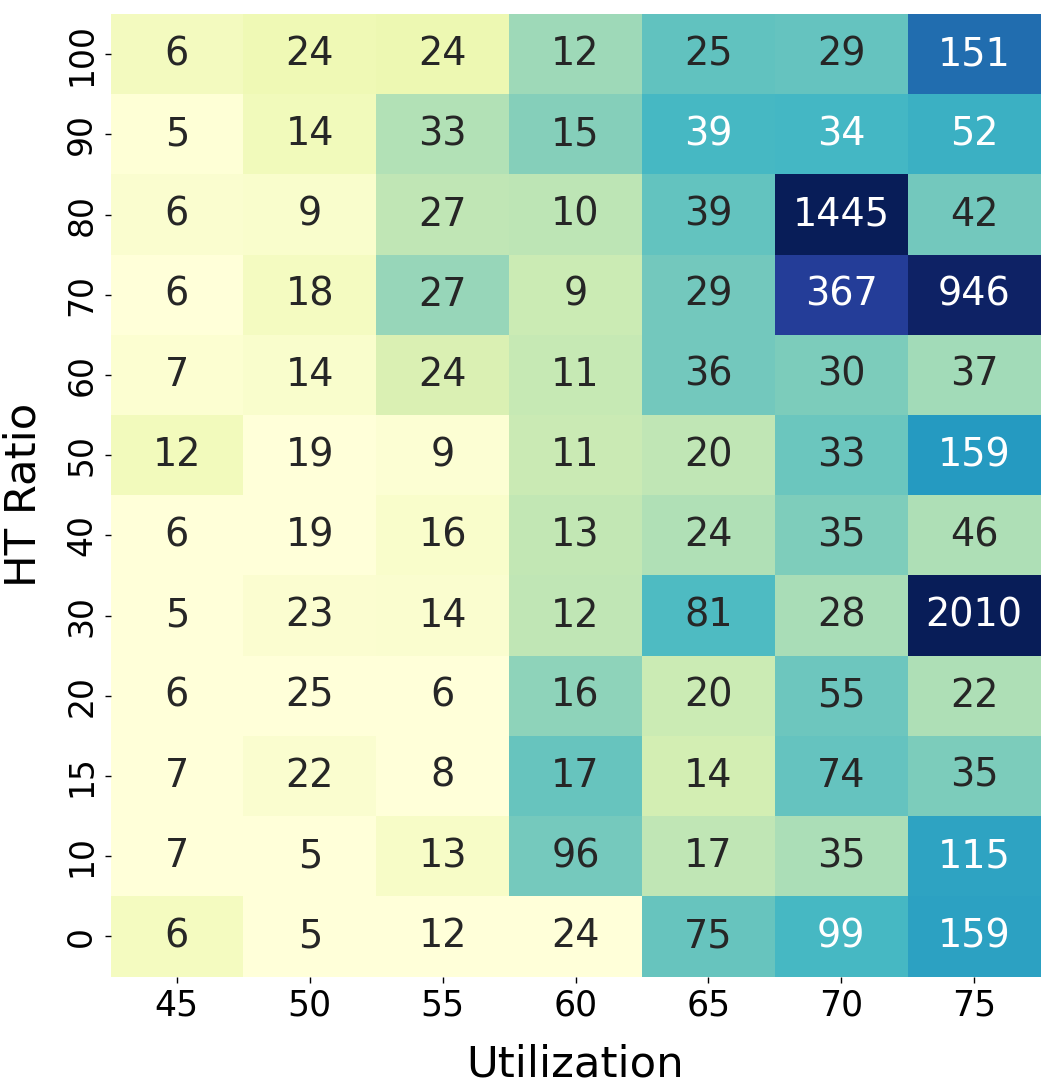}}
    \hspace{10pt}
    \subfigure[Extended SAG]{\includegraphics[scale=0.20]{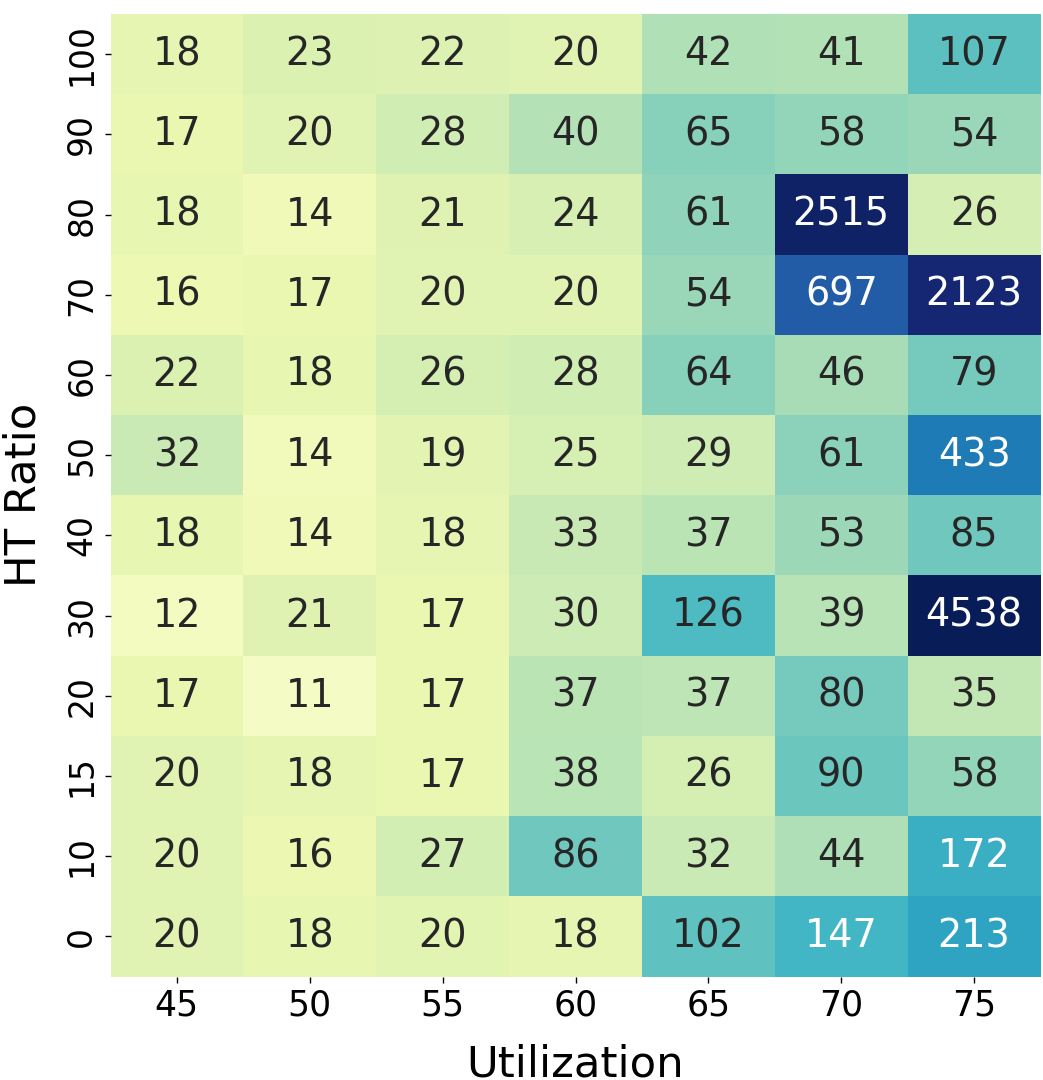}}
    \hspace{10pt}
    \subfigure[Hybrid SAG (proposed)]{\includegraphics[scale=0.20]{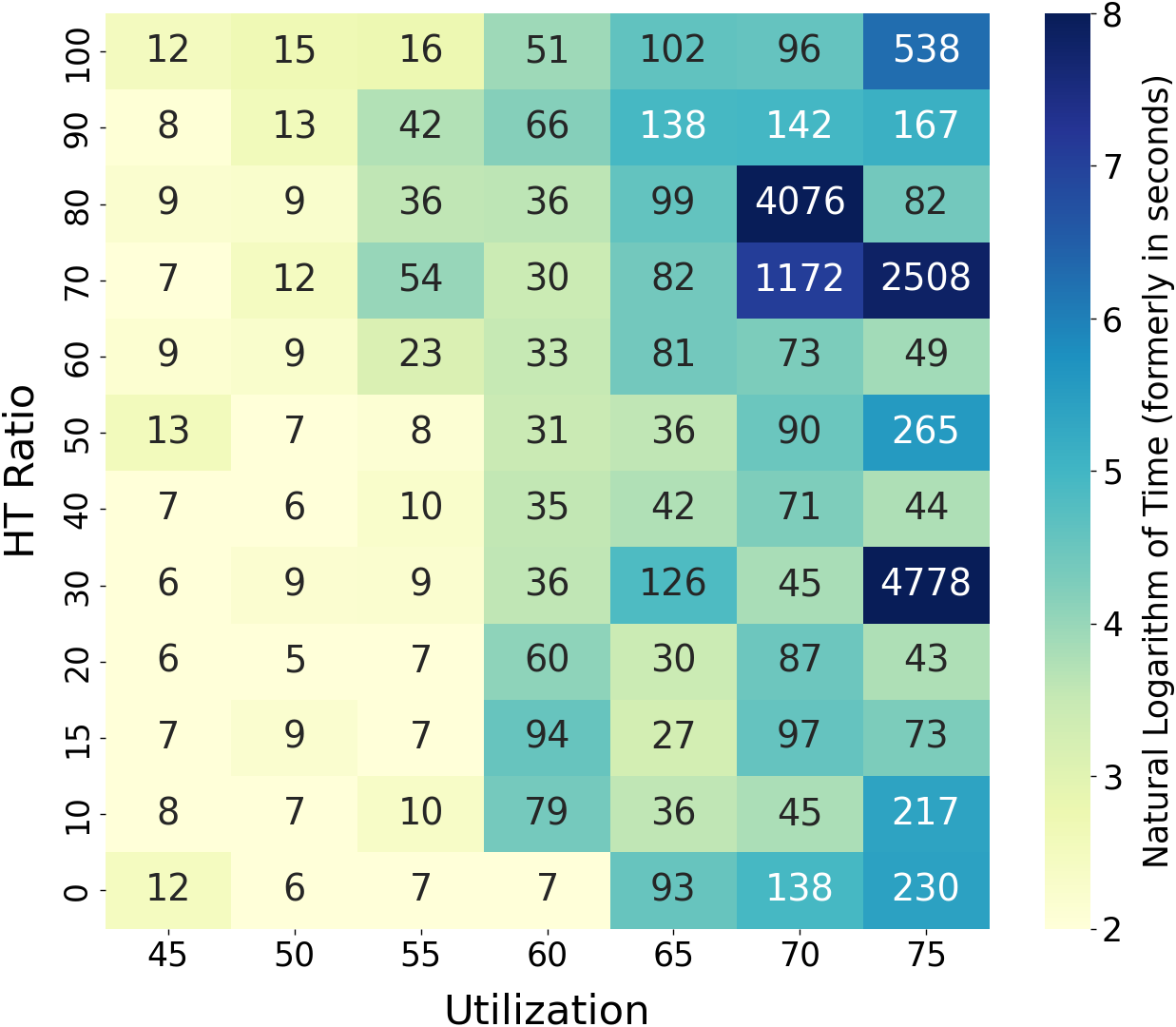}}
    \subfigure[Original SAG]{\includegraphics[scale=0.205]{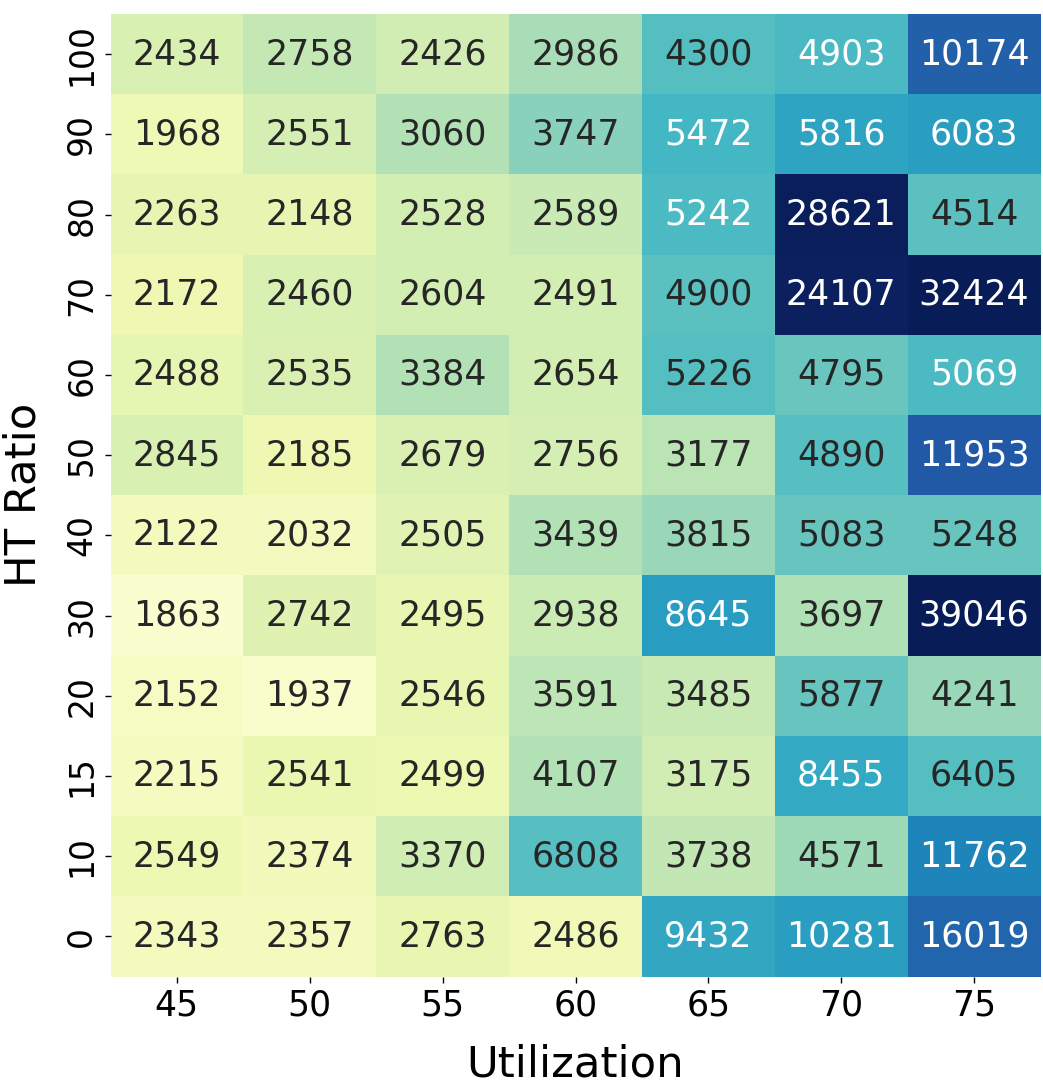}}
    \hspace{10pt}
    \subfigure[Extended SAG]{\includegraphics[scale=0.205]{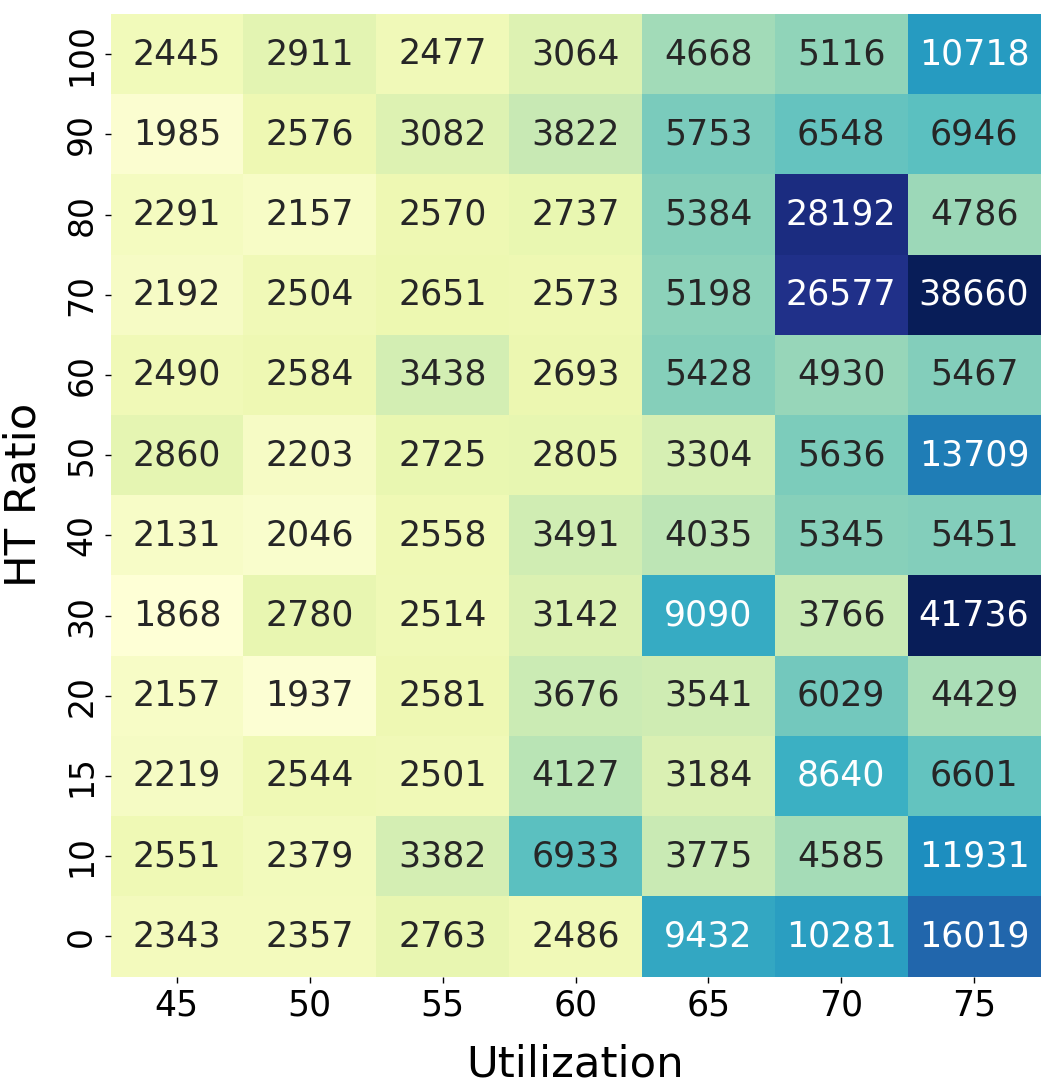}}
    \hspace{10pt}
    \subfigure[Hybrid SAG (proposed)]{\includegraphics[scale=0.205]{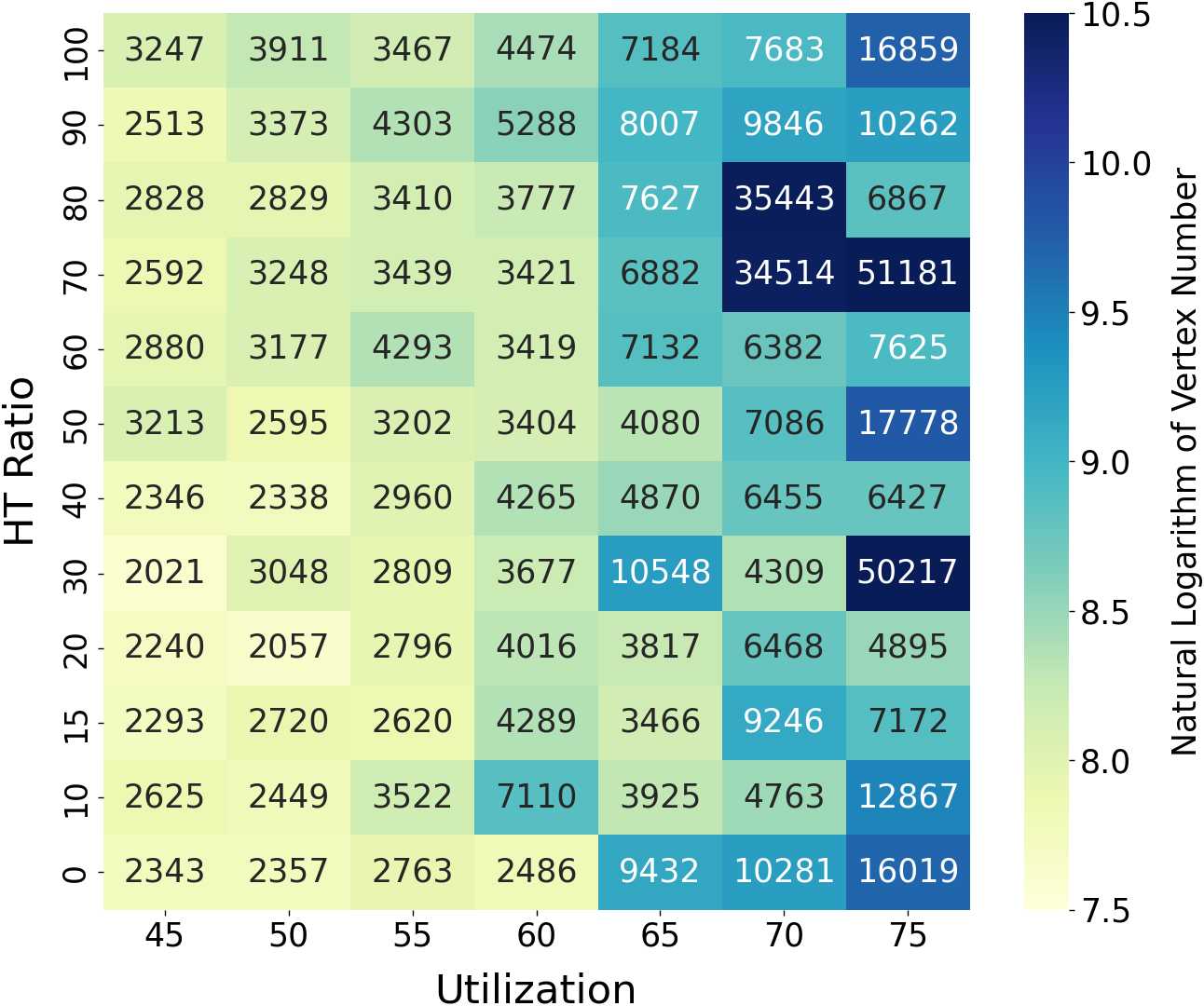}}
    \caption{Construction time and number of vertices for all cases using different construction algorithms.}
    \label{overhead_result}
\end{figure*}

\begin{table}[b]
    \caption{Ratio of the analyzed execution scenarios number to the actual execution scenarios number (common logarithm).}
    \centering
    \scalebox{0.82}{
        \begin{tabular}{|c|c|c|c|c|c|c|c|}
        \hline
        \textbf{Utilization (\%)} & 45     & 50     & 55     & 60     & 65     & 70     & 75     \\ \hline
        \textbf{Original}    & -17.88 & -16.19 & -17.94 & -18.39 & -19.67 & -15.92 & -19.92 \\ \hline
        \textbf{Extended}    & 13.79  & 17.92  & 24.34  & 29.77  & 36.28  & 32.96  & 46.54  \\ \hline
        \end{tabular}
    }
    \label{es_number_result}
\end{table}

We calculated the number of execution scenarios analyzed by different algorithms during construction using job sets containing 15\% HT jobs with varying utilization rates. The results show a dramatic difference as in Table \ref{es_number_result}. When the utilization was 60\%, the hybrid SAG considered exactly $10^{1198}$ actual possible execution scenarios, while the original SAG considered only $10^{-18.39}$ of them, and the extended SAG considered $10^{29.77}$ times them. Both the under-consideration of the original SAG and the over-consideration of the extended SAG have been far magnified as the number of jobs grows. The importance of being exact is evident here.

\begin{table}[b]
    \caption{Idle time required for safety.}
    \centering
    \scalebox{0.84}{
        \begin{tabular}{|c|c|c|c|c|c|c|c|}
        \hline
        \textbf{Utilization (\%)} & 45     & 50     & 55     & 60     & 65     & 70     & 75     \\ \hline
        \textbf{Idle Time}   & 300    & 344    & 431    & 545    & 646    & 584    & 850    \\ \hline
        \textbf{Efficiency}  & 97.0\% & 96.6\% & 95.7\% & 94.6\% & 93.5\% & 94.2\% & 91.5\% \\ \hline
        \end{tabular}
    }
    \label{idle_time_result}
\end{table}

As over-consideration of extended SAG leads to over-pessimism, especially when the original value of $C_i^{min}$ is relatively large, a job set could rarely be deemed schedulable. But for the original SAG, although it was over-optimistic because of its lack of consideration, there exists a straightforward way to keep the system running safely, i.e., \textit{idle-time insertion} \cite{howell1995non}. In the absence of an HT job, making the system idle during its originally assigned execution slots ensures safety. By doing so, all unconsidered scenarios can be manually transformed into considered ones. Nevertheless, inserting idle time can degrade the system's efficiency, as Table \ref{idle_time_result} shows. Moreover, the amount of idle time that needs to be inserted rises with system utilization. The higher the system utilization required, the less efficient the system can be.

\begin{table}[tb]
    \caption{Maximum width.}
    \centering
    \scalebox{0.9}{
        \begin{tabular}{|c|c|c|c|c|c|c|c|}
        \hline
        \textbf{Utilization (\%)} & 45 & 50  & 55 & 60  & 65 & 70  & 75  \\ \hline
        \textbf{Original}    & 35 & 80  & 42 & 314 & 60 & 383 & 89  \\ \hline
        \textbf{Extended}    & 35 & 80  & 42 & 314 & 60 & 383 & 90  \\ \hline
        \textbf{Hybrid}      & 38 & 106 & 42 & 314 & 82 & 384 & 116 \\ \hline
        \end{tabular}
    }
    \label{graph_width_result}
\end{table}

Another finding worth noting is that while there are numerous execution scenarios, the graph sizes of SAGs do not get that large. The depth of each SAG is exactly the size of the job set, i.e., 1000 in this case. Although the hybrid SAG creates extra vertices to maintain exactness, resulting in a larger width as shown in Table \ref{graph_width_result}, it is still well managed by the merging phase of the SAG construction algorithm.

\subsection{Scalability}

To more comprehensively evaluate the overhead of constructing SAGs and show their trends, we further generated 77 job sets for testing in addition to the 7 job sets with an HT rate of 15\%. The HT ratio and utilization of these job sets are taken crosswise from $\{45, 50, 55, 60, 65, 70, 75 \}$ and $\{0, 10, 20, 30, 40, 50, 60, 70, 80, 90, 100 \}$. A total of 252 SAG constructions were then performed on the 84 job sets using 3 algorithms, and the overhead results are shown in Fig. \ref{overhead_result}.

The time spent in the construction process for the original SAGs, the extended SAGs, and the hybrid SAGs are illustrated in Fig. \ref{overhead_result} (a), (b), and (c), respectively. With a relatively low HT ratio and utilization, it is even faster to build a hybrid SAG than to build an original SAG, while building an extended SAG is significantly slower. The speed advantage of hybrid SAGs gradually fades as the HT ratio and utilization rise. Analyzing the job set with an HT ratio of 30\% and utilization of 75\% was the most time-consuming for all three algorithms. They took 2010, 4538, and 4778 seconds, respectively. On this most challenging set of jobs, the time overhead of constructing a hybrid SAG is only 2.38 times that of an original one.

The number of vertices in the constructed SAGs is illustrated in Fig. \ref{overhead_result} (d), (e), and (f), respectively. In almost all cases, the size relations between the three kinds of SAGs are constant. An extended SAG is always not less than the corresponding original SAG, while the hybrid SAG is always no less than the extended one. A special case is the job set with an HT ratio of 80\% and utilization of 70\%, where the extended SAG is smaller than the original SAG. This could be attributed to the wider range of state reachable times in the extended SAG, making it easier for paths to \textit{match} each other and thus be merged.

Generally, the time and space overhead rises with the HT ratio and the utilization rate. More HT jobs may result in more possible orderings of job execution. The higher the utilization, the more jobs might be stacked together and the heavier the scheduling burden. However, the most difficult case to analyze is not the one with the highest HT ratio and utilization (topmost right corner), this should be due to the randomness of the job set generation.

In all cases, we observed that the number of vertices in the hybrid SAGs is at most 1.69 times higher than in the original SAGs, while the computing time is at most 5.42 times higher. On average, a hybrid SAG is only 1.82 times slower and 1.24 times larger than the original SAG. Since the scalability advantages (at least 3000 times faster than the UPPAAL-based exact RTAs) of the original SAG have been well evaluated in \cite{yalcinkaya2019exact, ranjha2022partial}, we consider the scalability of our proposed hybrid SAG to be sufficiently acceptable in these test cases.

\section{Conclusion and Future Work}

In this work, we extended the SAG framework to support HT jobs through the development of hybrid SAG, enabling a more flexible and widely applicable system model. Hybrid SAG provides a robust solution for modeling complex job behaviors, including release time jitter, execution time variation, and absence, while maintaining exactness. Experimental results show the scalability and effectiveness of our approach. By releasing SAGkit as an open-source Python toolkit, we aim to empower the research community to explore new directions in fault-tolerant distributed control, RTA, and SAG, thus advancing the field and promoting innovation.

In the future, we plan to enhance hybrid SAG to support general non-consecutive execution times cases. The absence of HT jobs is, in fact, a specific instance of this scenario. Through the extended SAG, we have demonstrated that including unreachable execution scenarios can lead to overly pessimistic results. Thus, robust support for non-consecutive execution times is essential for SAG to maintain exactness in the analysis of more complex job sets.

Furthermore, we aim to expand SAGkit’s capabilities to accommodate scenarios in which jobs have dependencies on previous executions, allowing for a more nuanced analysis of state-dependent job behaviors. In many practical systems, the behavior or timing of a job is influenced by prior executions, which creates interdependencies that affect the exactness of RTA. By integrating support for dependencies, SAGkit would enable a finer-grained representation of such systems, reducing the conservatism of traditional approaches that often overestimate requirements.

\bibliographystyle{IEEEtran}
\bibliography{SAGkit_REF}

\end{document}